\documentclass[sigconf]{acmart}
\AtBeginDocument{%
  }

\setcopyright{rightsretained}
\copyrightyear{2025}
\acmYear{2025}
\acmConference{SIGGRAPH Posters '25}{August 10-14, 2025}{Vancouver, BC, Canada}\acmBooktitle{Special Interest Group on Computer Graphics and Interactive Techniques Conference Posters (SIGGRAPH Posters '25), August 10-14, 2025}\acmDOI{10.1145/3721250.3743033}
\acmISBN{979-8-4007-1549-5/2025/08}

\usepackage{graphicx}
\usepackage{float}
\usepackage{animate}
\usepackage{subcaption}
\usepackage{wrapfig}

\begin{document}

\title{Sketch-based Fluid Video Generation Using Motion-Guided Diffusion Models in Still Landscape Images}

\author{Hao Jin}
\orcid{0009-0005-6722-9322}
\affiliation{%
  \institution{Japan Advanced Institute of Science and Technology}
  \city{Nomi}
  \state{Ishikawa}
  \country{Japan}
}

\author{Haoran Xie}
\orcid{0000-0002-6926-3082}
\affiliation{%
  \institution{Japan Advanced Institute of Science and Technology}
  \city{Nomi}
  \state{Ishikawa}
  \country{Japan}
}

\renewcommand{\shortauthors}{Jin et al.}


\begin{teaserfigure}
\centering
  \includegraphics[width=0.99\linewidth]{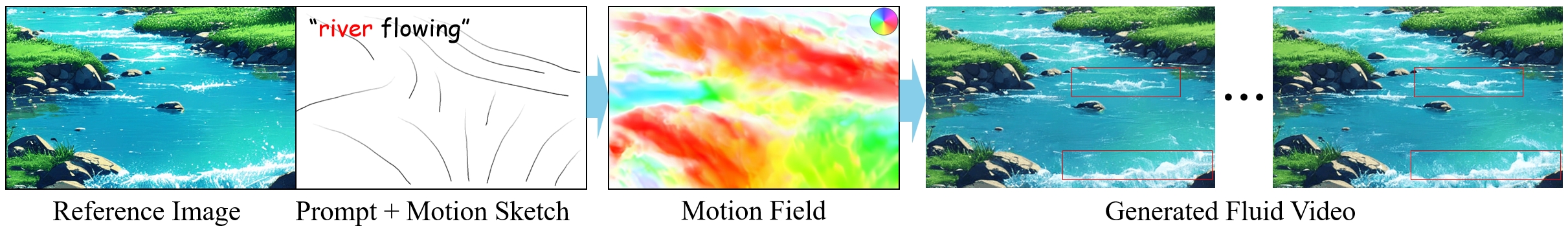}
  \caption{Given a reference landscape image and hand-drawn motion sketches, our framework automatically animates still fluid through two main stages: motion field generation and motion-guided video synthesis. }
  \label{fig:teaser}
\end{teaserfigure}


\maketitle

\section{Introduction}
Integrating motion into static images not only enhances visual expressiveness but also creates a sense of immersion and temporal depth, establishing it as a longstanding and impactful theme in artistic expression. Fluid elements such as waterfall, river, and oceans are common features in landscape, but their complex dynamic characteristics pose significant challenges in modeling and controlling their motion within visual computing. Physics-based methods are often used in fluid animation to track particle movement. However, they are easily affected by boundary conditions. Recently, latent diffusion models have been applied to video generation tasks, demonstrating impressive capabilities in producing high-quality and temporally coherent results. However, it is challenging for the existing methods to animate fluid smooth and temporally consistent motion. To solve these issues, this paper introduces a framework for generating landscape videos by animating fluid in still images under the guidance of motion sketches as shown in Figure \ref{fig:teaser}. We propose a finetuned conditional latent diffusion model for generating motion field from user-provided sketches, which are subsequently integrated into a latent video diffusion model via a motion adapter to precisely control the fluid movement.

\section{Methods}
\begin{figure}[t]
  \includegraphics[width=0.99\linewidth]{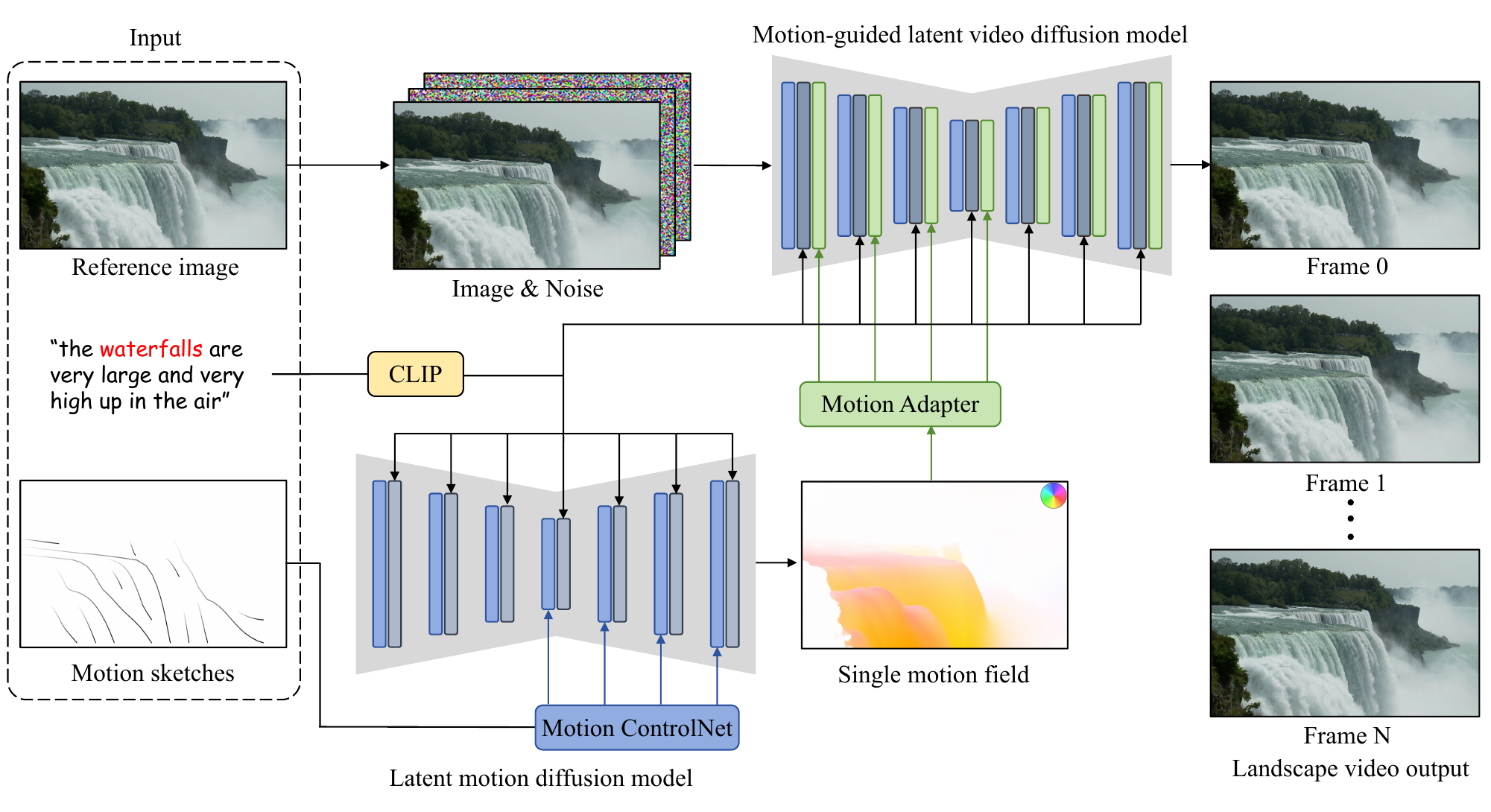}
  \caption{The workflow of our proposed sketch-based video fluid generation framework.}
  \label{fig:framework}
\end{figure}

Figure \ref{fig:framework} illustrates the workflow of our proposed framework for fluid video generation guided by hand-drawn sketches. 

\begin{figure}[t]
  \includegraphics[width=0.99\linewidth]{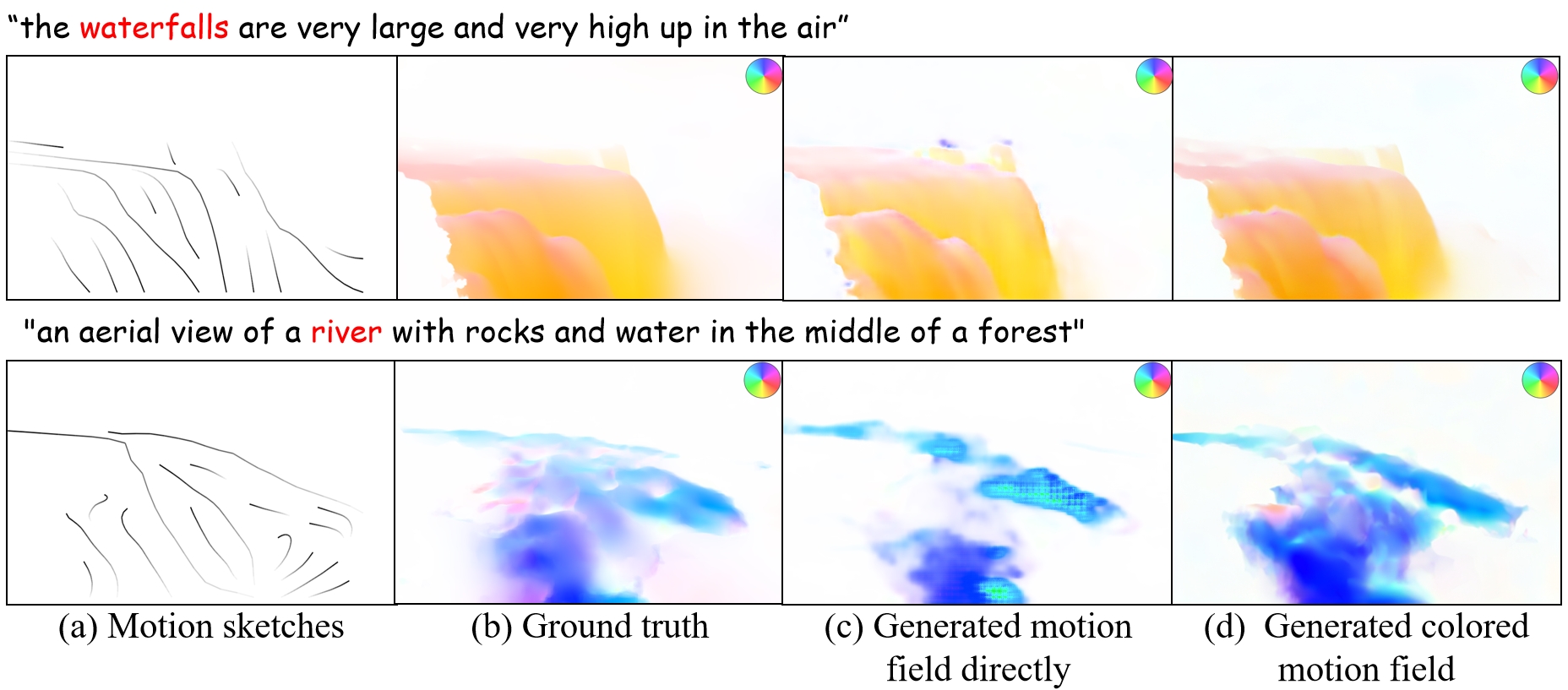}
  \caption{Comparison between the generated motion fields with the colored representation.}
  \label{fig:motion}
\end{figure}

\textbf{\textit{Motion field estimation}}. Compared with text prompts and trajectories, sketches serve as a more intuitive medium for user expression, allowing flexible manipulation of motion. Inspired by Sketch2Cinemagraph \cite{jin2024sketch}, this paper employs a ControlNet to generate motion fields conditioned on user-provided motion sketches with text. The motion sketches are presented as gradient gray lines to align with the streamlines extracted from the ground truth motion fields in the landscape dataset \cite{holynski2021animating}. Different from previous work \cite{holynski2021animating}, this paper chooses to estimate the colored representation of motion fields, instead of directly predicting the motion fields themselves. By reformulating motion field generation as an image generation problem, this approach leads to substantial reductions in both training time and computational resource requirements. Figure \ref{fig:motion} shows the colored motion fields generated by our latent motion diffusion model. In comparison with directly generated motion fields, the colored versions retain finer details, which more effectively support downstream video generation guidance.

\textbf{\textit{Motion-guided fluid generation}}. We leverage the Latent Video Diffusion Model (LVDM) to animate fluid dynamics within still landscape images. To turn a text-to-video backbone into an image-to-video one, we add a reference image’s CLIP embedding to the starting noise. This preserves the content of the first frame and promotes spatiotemporal coherence across the generated video sequence. To incorporate motion guidance, we introduce a motion adapter that encodes the colored representation of the motion field into the latent space. The encoded motion features are subsequently injected into the temporal transformer layers of the denoising 3D U-Net, enabling the model to synthesize temporally consistent and motion-aware video content. Unlike the previous approach \cite{wang2024motionctrl} that relies on a sequence of motion fields as conditions, our method utilizes a single motion field to guide the fluid dynamics. This is motivated by the observation that real-world fluid motion typically exhibits strong temporal stability, allowing a single motion field to capture its primary dynamic characteristics effectively. Moreover, this approach significantly reduces training cost and computational overhead.

\textbf{\textit{Anime-style fluid generation}}. Our motion-guided LVDM is trained on real-world landscape datasets \cite{holynski2021animating}. Benefiting from the style transfer capabilities of diffusion models, our framework generalizes beyond the training domain and accepts anime-style landscape images as reference inputs. These anime-style images can be made with text-to-image models or by applying style transfer to real landscape photos. Guided by motion sketches, our framework produces fluid videos that keep realistic movement and the anime visual aesthetic. Figure \ref{fig:results} showcases examples of anime-style fluid videos generated by our framework.

\section{Results}
Table \ref{tab: compare} shows a comparison between our framework and Stable Video Diffusion (SVD) \cite{blattmann2023stable}. The quality of the generated fluid was assessed by computing the FVD between the synthesized fluid video and the corresponding ground truth video. Each video has 25 frames at a resolution of $512\times320$. The results show that our framework generates more realistic fluid dynamics in landscape videos. Details are shown in the supplementary video. 

Figure \ref{fig:results} shows that the generated video frames across various fluid scenarios highlight our framework’s capability to handle complex motion patterns. In addition, the anime-style results show that the movement of generated fluid not only adheres closely to the user-provided sketch guidance but also maintains a consistent anime style throughout the entire frame sequence.

\begin{figure}[t]
	\centering
        
        \begin{subfigure}{0.28\linewidth}
		\includegraphics[width=\textwidth]{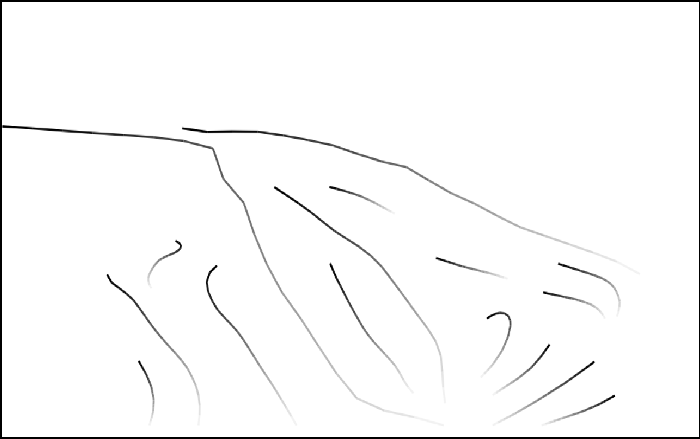}
	\end{subfigure}
        \begin{subfigure}{0.28\linewidth}
		\includegraphics[width=\textwidth]{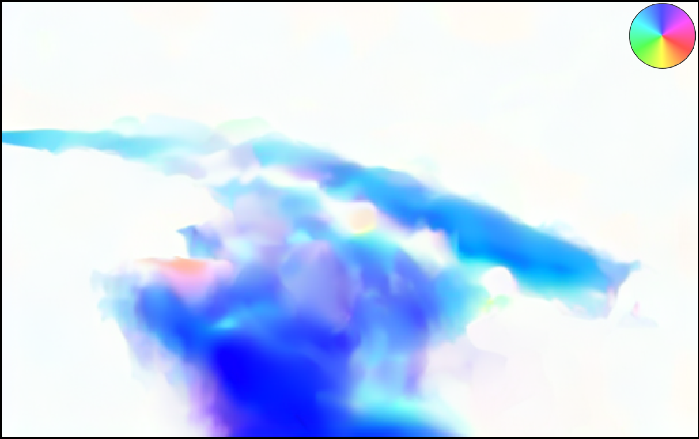}
	\end{subfigure}
	\begin{subfigure}{0.28\linewidth}
		\animategraphics[autoplay,loop, width=\textwidth]{8}{images/gif/2/frames/frame}{0}{24}
	\end{subfigure}

	\begin{subfigure}{0.28\linewidth}
		\includegraphics[width=\textwidth]{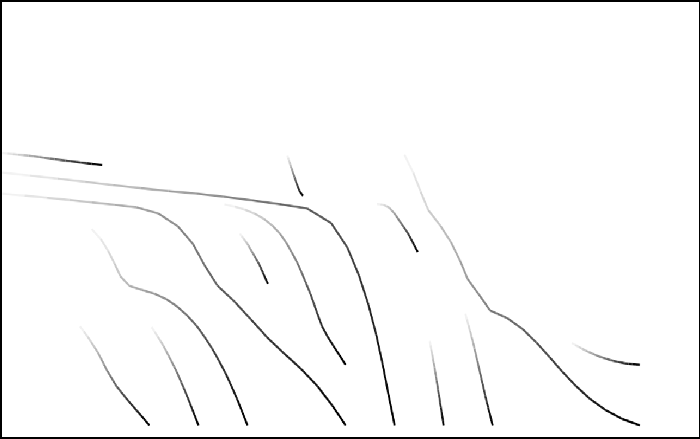}
	\end{subfigure}
        \begin{subfigure}{0.28\linewidth}
		\includegraphics[width=\textwidth]{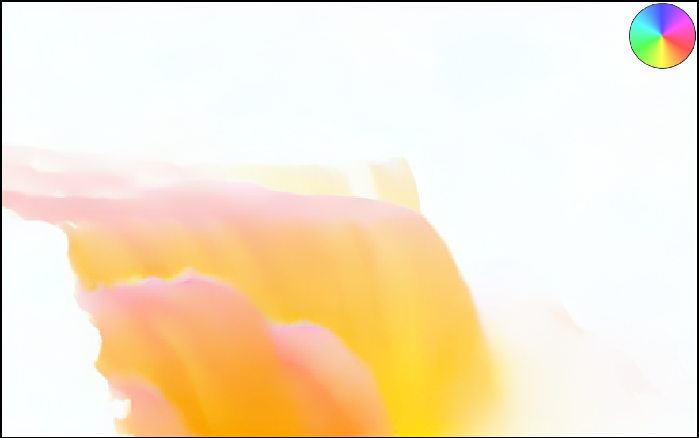}
	\end{subfigure}
	\begin{subfigure}{0.28\linewidth}
		\animategraphics[autoplay,loop, width=\textwidth]{8}{images/gif/0/frames/frame_}{0}{24}
	\end{subfigure}

        \begin{subfigure}{0.28\linewidth}
		\includegraphics[width=\textwidth]{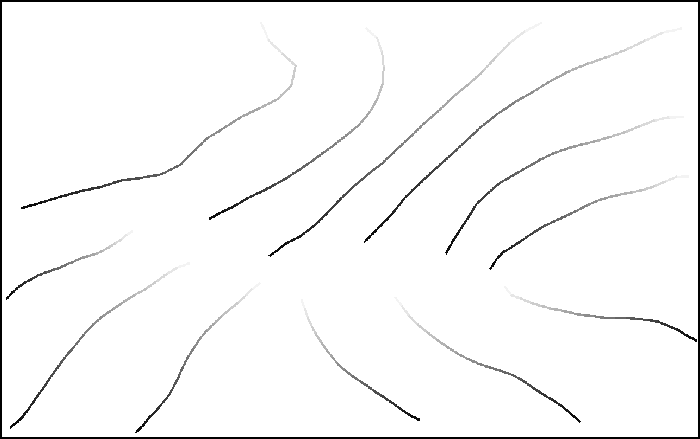}
            \caption{Motion sketches}
	\end{subfigure}
        \begin{subfigure}{0.28\linewidth}
		\includegraphics[width=\textwidth]{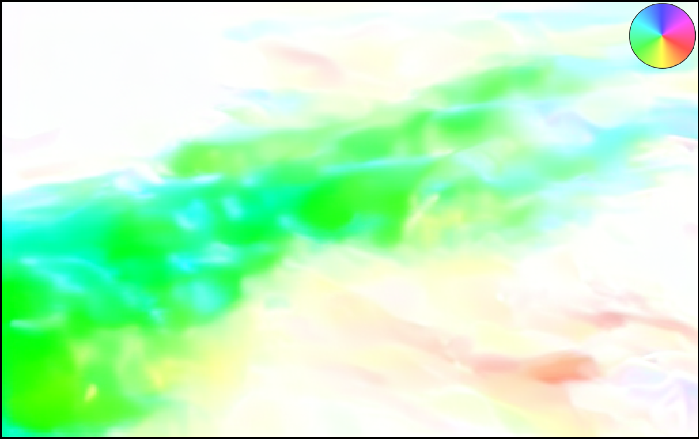}
        \caption{Motion fields}
	\end{subfigure}
	\begin{subfigure}{0.28\linewidth}
		\animategraphics[autoplay,loop, width=\textwidth]{8}{images/gif/7/frames/frame}{0}{24}
        \caption{Fluid videos}
	\end{subfigure}

	\captionof{figure}{Examples of fluid video generation (c) guided by motion fields (b) produced from motion sketches (a). (\textbf{These videos are embedded and better viewed using Adobe Reader.})}
	\label{fig:results}
\end{figure}

\begin{table}[!h]
\begin{center}
\caption{Comparison results with previous work.}
\begin{tabular}{c|c|c}
\hline
Method        & LPIPS↓          & PSNR↑            \\ \hline
SVD           & 0.3508         & 13.2409          \\ \hline
\textbf{Ours} & \textbf{0.3049} & \textbf{15.6186} \\ \hline
\end{tabular}
\label{tab: compare}
\end{center}
\end{table}

\section{Conclusion}
This paper introduced a novel framework for generating fluid videos in still landscape images guided by hand-drawn motion sketches. The proposed method is capable of synthesizing fluids with realistic dynamics in both naturalistic and anime-style landscape scenes. The fluid motion can be intuitively guided by the user, making the approach well-suited for creative video production.

\begin{acks}
This work is funded by New Energy and Industrial Technology Development Organization (NEDO) JPNP20017, JSPS KAKENHI Grant Numbers 23K18514 and 25K00154.
\end{acks}

\bibliographystyle{ACM-Reference-Format}
\bibliography{sample-base}


\end{document}